\documentstyle[12pt]{article}
\begin{document}
%
%
\begin{center}
{\bf Quantum Computation as a Dynamical Process}\\ \ \\
G.P. Berman$^1$, G.D. Doolen$^1$, and V.I. Tsifrinovich$^2$\\ \ \\
\end{center}
$^1$Theoretical Division and CNLS, Los Alamos National Laboratory, \\
Los Alamos, New Mexico 87545\\
$^2$Department of Applied Mathematics and Physics, Polytechnic University,\\
Six Metrotech Center, Brooklyn NY 11201\\ \ \\
\begin{center}
{\bf ABSTRACT}
\end{center}
In this paper, we discuss the dynamical issues of quantum computation. We demonstrate that fast wave function oscillations can affect the performance of Shor's quantum algorithm by destroying required quantum interference. We also show that this destructive effect can be routinely avoided by using resonant-pulse techniques. We discuss the dynamics of resonant pulse implementations of quantum logic gates in Ising spin systems. We also discuss the influence of non-resonant excitations. We calculate the range of parameters where undesirable non-resonant effects can be minimized. Finally, we describe the ``$2\pi k$-method'' which avoids the detrimental deflection of non-resonant qubits. 
\newpage
\quad\\
{\bf I. Shor's quantum algorithm -- the simplest example}\\ \ \\

Successful design of quantum computers requires a thorough understanding of 
the time evolution of quantum qubits. The current interest in quantum computation was stimulated  by Shor \cite{s1} who invented the quantum algorithm for prime factorization of integers. Shor's algorithm has three main steps.  (See (7) through (11) below.) We discuss here the simplest example of Shor's algorithm and use it in the next section to demonstrate the role of fast oscillations. e
Suppose we want to factor the smallest composite number, 4. According to Shor's method, we choose the only coprime number, 3. A quantum computer will then compute the periodic function,
$$
y(x)=3^x~(mod~4),\eqno(1)
$$
which is the remainder after division of $3^x$ by 4. This function is,
$$
y(0)=1,\quad y(1)=3,\quad y(2)=1,\quad y(3)=3,...\eqno(2)
$$
Then, the quantum computer must find the period of this function, $T=2$, in our case. The factor of 4 can be found as the greatest common divisor (GCD) of two numbers: $(z+1)$ and 4 or $(z-1)$ and 4, where $z=3^{T/2}$. In our case, 
$$
GCD(z-1,4)=GCD(2,4)=2,\eqno(3)
$$
gives the desired factor of 4.

The simplest quantum computer which computes the function (1) and finds its period, $T$, has 4 quantum bits (qubits): the two left-most qubits represent the number ``x'', and  the two right-most qubits represent ``y'',
$$
x=2m_1+m_0,\quad y=2n_1+n_0,\quad m_i,n_i=0,1,~(i=0,1).\eqno(4)
$$
As an example, the values $x=1$ and $y=3$ will be represented as,
$$
|m_1m_0,n_1n_0\rangle=|01,11\rangle.\eqno(5)
$$
We use the Dirac notation for the quantum states: $|0\rangle$ for the ground state and $|1\rangle$ for the excited state.

Assume that initially all four qubits are in their ground states. The wave function of the system is,
$$
\Psi_0=|00,00\rangle.\eqno(6)
$$
Following Shor's idea, the quantum computer must carry out three main unitary transformations:\\
1) It creates a superposition of all possible values of ``x'',
$$
\Psi_1={{1}\over{2}}(|00,00\rangle+|01,00\rangle+|10,00\rangle+|11,00\rangle).\eqno(7)
$$
2) It computes the function, $y(x)$, in (1) using a digital algorithm, for all values of ``x'' simultaneously,
$$
\Psi_2={{1}\over{2}}(|00,01\rangle+|01,11\rangle+|10,01\rangle+|11,11\rangle).\eqno(8)
$$
3) It performs a discrete Fourier transform (DFT) for each value of ``x'',
$$
|x\rangle\rightarrow{{1}\over{2}}\sum_{k=0}^3e^{2\pi ikx/4}|k\rangle.\eqno(9)
$$
For example,
$$
|00,01\rangle\rightarrow{{1}\over{2}}(|00,01\rangle+|01,01\rangle+|10,01\rangle+|11,01\rangle).\eqno(10)
$$
As a result, 
$$
\Psi_3={{1}\over{2}}(|00,01\rangle+|00,11\rangle+|10,01\rangle-|10,11\rangle).\eqno(11)
$$
Measuring the value of ``x'' one gets either: $x=x_1=0$ or $x=x_2=2$. The ratio  $D/x_2$ (where $D=4$ is the number of all possible values of ``x'') is the period of the periodic function $y(x)$: $T=D/x_2=2$ \cite{ek}.\\ \ \\
{\bf II. Fast oscillations}\\ \ \\

Note that Shor's algorithm is described above as a sequence of {\it instantaneous} unitary transformations. What happens if each unitary transformation takes a finite time? 
In time, $t$, in quantum mechanics, each stationary state, $|k\rangle$, of the quantum system acquires its own phase, $-E_kt/\hbar$. This phase is usually responsible for fast oscillations of the wave function. The slow dynamics is caused by small perturbations which are responsible for transitions between the energy levels. 
Generally it is not obvious that the DFT will provide the desired constructive and destructive interferences if one takes into account the fast oscillations of wave function during the finite time intervals of quantum transformations and possible time delays between these transformations.

Suppose, for simplicity, that there are finite time delays between three main unitary transformations described in section I, while the time duration of each transformation is, as before, infinitely small. Then, immediately before the second transformation, each state in the superposition (7) will possess the ``natural'' phase factor,
$$
\Psi_1={{1}\over{2}}\Bigg(|00,00\rangle e^{-iE_{00}\tau_1}+ |01,00\rangle e^{-iE_{10}\tau_1}+|10,00\rangle e^{-iE_{20}\tau_1}+ |11,00\rangle e^{-iE_{30}\tau_1}\Bigg),\eqno(12)
$$
where we put $\hbar=1$; $E_{ab}$ is the energy corresponding to the state: $x=a$, $y=b$; $\tau_1$ is the time delay between the first two transformations. The second transformation generates new states, and the ``natural'' connection between the state and the phases vanishes. Just before the third transformation we have,
$$
\Psi_2=\{|00,01\rangle\exp(-iE_{00}\tau_1-iE_{01}\tau_2)+
         |01,11\rangle\exp(-iE_{10}\tau_1-iE_{13}\tau_2)+\eqno(13)
$$
$$
|10,01\rangle\exp(-iE_{20}\tau_1-iE_{21}\tau_2)+
|11,11\rangle\exp(-iE_{30}\tau_1-iE_{33}\tau_2)\},
$$
where $\tau_2$ is the time delay between the second and the third unitary transformations. It is easy to check that the desired interference does not occur after the DFT. For example, the state $(1/2)|00,01\rangle$ in the superposition (11) is described now by two terms,
$$
{{1}\over{4}}|00,01\rangle[\exp(-iE_{00}\tau_1-iE_{01}\tau_2)+\exp(-iE_{20}\tau_1-iE_{21}\tau_2)].\eqno(14)
$$
The state $|01,01\rangle$ (which vanishes if $\tau_1=\tau_2=0$) now survives, and has the form,
$$
{{1}\over{4}}|01,01\rangle[\exp(-iE_{00}\tau_1-iE_{01}\tau_2)-\exp(-iE_{20}\tau_1-iE_{21}\tau_2)].\eqno(15)
$$
The reason for breaking Shor's algorithm in this example is that the ``like'' terms corresponding to the same state, carry the ``history'' of their origination. For example, the term,
$$
{{1}\over{4}}|00,01\rangle\exp(-iE_{00}\tau_1-iE_{01}\tau_2),\eqno(16)
$$
in (14) originated from the transformations,
$$
|00,00\rangle\rightarrow|00,00\rangle\stackrel{\tau_1}{\rightarrow}|00,01\rangle
\stackrel{\tau_2}{\rightarrow}|00,01\rangle.\eqno(17)
$$
The term,
$$
{{1}\over{4}}|00,01\rangle\exp(-iE_{20}\tau_1-iE_{21}\tau_2),\eqno(18)
$$
originated from the transformation,
$$
|00,00\rangle\rightarrow|10,00\rangle\stackrel{\tau_1}{\rightarrow}|10,01\rangle
\stackrel{\tau_2}{\rightarrow}|00,01\rangle.\eqno(19)
$$
One can see that the ``phase memory'' includes the energies connected with the process of formation of the corresponding terms. The question arises:  ``How can one carry out a quantum computation with finite-time intervals between pulses and not destroy the desired interferences?''

The natural way to solve this problem is to ``produce'' the ``natural'' phase, $-E_{ab}\Delta t$, when the term corresponding to $x=a$ and $y=b$ is generated in the process of quantum computation. (Here, $\Delta t$, is the time interval between the beginning of the first transformation and the end of the transformation which generates this state, $|x=a,y=b\rangle$.) We now discuss how this can be done routinely for a resonant technique such as NMR.\\ \ \\
{\bf III. Resonant pulse implementation of quantum unitary transformations}\\ \ \\

Recently, two basic approaches have been used for implementing unitary transformations for quantum computation. The most popular approach originated from the paper by Lloyd who proposed using resonant electromagnetic pulses to perform any required transformation \cite{l1}. These pulses drive the quantum transitions between the corresponding energy levels in a system of weakly interacting particles. Following Lloyd's idea, we proposed in \cite{b1} using Ising spin systems for quantum computation. For such systems, the ``digital states'' corresponding to specific values of $x$ and $y$, are true eigenstates of the Hamiltonian including interaction between spins. This idea has been implemented in nuclear spin systems in liquids, which are closely 
approximated by Ising systems \cite{g1,c1}. It was found in \cite{g1,c1} that quantum computation can be carried out on a statistical ensemble of such systems at room temperature. There are also a few other important ``resonant'' proposals including a well-known idea of the ion trap quantum computer \cite{c2}.

The second basic approach does not require resonant pulses. The earliest proposals of this kind were reviewed in \cite{land}, and the most recent idea was suggested in \cite{d1}. Here, we are considering only the ``resonant'' proposals. In this section, we discuss how the ``natural'' phase,
$-E_{ab}\Delta t$, can be routinely generated in a quantum computer using  resonant pulses.

We assume that there is a required sequence of resonant pulses to produce a quantum computation. Each pulse induces a selective transition between two definite states of the system. We denote the state of the whole system by a single index, e.g. $|n\rangle$. The transition between two states, $|k\rangle$ and $|n\rangle$, can be described by the following effective Hamiltonian:
$$
{\cal H}=E_k|k\rangle\langle k|+E_n|n\rangle\langle n|-{{\Omega}\over{2}}\Bigg[
e^{i(\omega t+\varphi)}|k\rangle\langle n|+
e^{-i(\omega t+\varphi)}|n\rangle\langle k|\Bigg].\eqno(20)
$$
Here $E_k$ and $E_n$ are the energies of the corresponding states; $\Omega$ is the Rabi frequency; $\omega=\omega_{nk}\equiv E_n-E_k$ is the resonant frequency, and $\varphi$ is the phase of the resonant pulse. We assume that one imposes  the following  resonant fields,
$$
h_+=h\exp(-i\omega_{nk}t),\eqno(21)
$$
for each resonant transition required for the quantum computation. 
Suppose that the $j$-th pulse induces a transition from  state $|k\rangle$ to state $|n\rangle$. We assume that at the beginning of the pulse, the amplitude, $C_k$, of the state $|k\rangle$ has the ``natural'' phase factor,
$$
\exp(-iE_kt_{j-1}),\quad t_{j-1}=\tau_1+\tau_2+...+\tau_{j-1},\eqno(22)
$$
where $\tau_k$ is the duration of the $k$-th pulse. We also assume that there is no delay between two consecutive pulses. (This last assumption is not crucial for our consideration.) The Schr\"odinger equation for the amplitudes $C_k$ and $C_n$ has the form,
$$
i\dot C_k=E_kC_k-{{1}\over{2}}\Omega e^{i(\omega t+\varphi)}C_n,\quad i\dot C_n=E_nC_n-{{1}\over{2}}\Omega e^{-i(\omega t+\varphi)}C_k.\eqno(23)
$$
At the end of the $j$-th pulse ($t=t_j=t_{j-1}+\tau_j$), we have,
$$
C_k(t_j)=\exp[i(E_kt_{j-1}-E_kt_j)]\cos\alpha_jC_k(t_{j-1}),\quad \alpha_j=\Omega\tau_j/2,\eqno(24)
$$
$$
C_n(t_j)=\exp[i(\pi/2-\varphi+E_kt_{j-1}-E_nt_j)]\sin\alpha_jC_k(t_{j-1}).
$$
Using (22), we can rewrite these expressions in the form,
$$
C_k(t_j)=|C_k(t_{j-1})|\exp(-iE_kt_j)\cos\alpha_j,\eqno(25)
$$
$$
C_n(t_j)=|C_k(t_{j-1})|\sin\alpha_j\exp[(i(\pi/2-\varphi)]\exp(-iE_nt_j).
$$
One can see from (25) that the new state, $|n\rangle$, has automatically acquired the ``natural'' phase, $-E_nt_j$. The factor, $\exp[i(\pi/2-\varphi)]$, in (25) describes the ``standard'' phase shift which can be eliminated by choosing: $\varphi=\pi/2$.\\ \ \\
{\bf IV. Non-resonant action of resonant pulses}\\ \ \\
A major obstacle for the resonant pulse implementation  of quantum computation is the non-resonant effect of resonant pulses on some qubits. This action can be completely avoided only if each qubit is isolated from all other qubits. This is not possible or desirable in reality. One can minimize non-resonant effects by providing a significant difference between the resonant frequencies.
  We shall now discuss this possibility using the example of a two-qubit quantum Control-Not gate which is described by the operator,
$$
CN=|00\rangle\langle 00|+|01\rangle\langle 01|+|10\rangle
\langle 11|+|11\rangle\langle 10|.\eqno(26)
$$
This gate changes the value of the right (target) qubit if the left (control) qubit is in the excited state. The importance of the CN gate became evident when it was shown that any unitary transformation can be implemented using combinations of one-qubit rotations and two-qubit CN gates \cite{b2}. A single $\pi$-pulse implementation of the CN gate was first suggested for 
coupled quantum dots \cite{b3}. We have studied this implementation using 
Ising spin systems \cite{b4}-\cite{b9}. We investigated the dynamics of a single $\pi$-pulse CN gate in the reference frame rotating with frequency, $\omega$, of the circularly polarized magnetic field, which induces a resonant transition
of a target spin, if the control spin is in the excited state. The effective Hamiltonian in the rotating frame has the form \cite{b9}:
$$
{\cal H}=-\sum_k\Bigg[(\omega_k-\omega)I^z_k+2\sum_{n\not=k}J_{kn}I^z_kI^z_n+\Omega_kI^x_k\Bigg],\eqno(27)
$$
where $\omega_k$ is the resonant frequency of a single $k$-th spin and $J_{kn}$ are the Ising interaction constants. The utility of this Hamiltonian for 
liquid NMR experiments was tested and confirmed in \cite{c3}. We have shown \cite{b6,b9} that a single pulse performs the CN gate operation,
$$
{\cal CN}=|00\rangle\langle 00|+|01\rangle\langle 01|+i|10\rangle
\langle 11|+i|11\rangle\langle 10|,\eqno(28)
$$
accompanied by a $\pi/2$ phase shift for the transformation of the target qubit. Numerical calculations for a well-separated frequencies,
$$
\omega_1=500,\quad \omega_2=100,\quad J=5,\quad\omega=\omega_2-J=95,\quad \Omega_1=0.5, \quad \Omega_2=0.1,\eqno(29)
$$
confirmed that non-resonant effects are negligible for the single-pulse  implementation of the CN gate \cite{b4,b9}. For example, the initial superpositional state,
$$
\Psi_0=\sqrt{0.3}|00\rangle+\sqrt{0.2}|01\rangle+{{1}\over{\sqrt{3}}}|10\rangle+{{1}\over{\sqrt{6}}}|11\rangle,\eqno(30)
$$
transforms as a result of the ${\cal CN}$ gate (28) to the state (See Fig 25.2 in \cite{b9}.),
$$
\Psi_1=\sqrt{0.3}|00\rangle+\sqrt{0.2}|01\rangle+(i/\sqrt{6})|10\rangle+(i/\sqrt{3})|11\rangle.\eqno(31)
$$
We have also investigated the dynamics of the CN gate for an ensemble of four-spin molecules at room temperature: $k_BT>>\hbar \omega_k$ \cite{b7}-\cite{b9}. Following the idea suggested in \cite{g1}, we implemented quantum transformations for four active states, $|00ij\rangle$, ($i,j=0,1$) which can be prepared initially in the ground state, $|0000\rangle$. The corresponding density matrix has the form,
$$
\rho(t)=E/16+\rho_\Delta(t),\eqno(32)
$$
where $\rho_\Delta$ is the following deviation density matrix:
$$
\rho_\Delta(t)={{\sum_{k=0}^3\hbar\omega_k}\over{32k_BT}}\Bigg[
\sum_{n,p=0}^{3}r_{n,p}(t)|n\rangle\langle p|+
\sum_{n,p}b_{n,p}(t)|n\rangle\langle p|\Bigg].\eqno(33)
$$
The first sum in (33) describes the dynamics of the 16 density matrix elements, $r_{n,p}(t)$, corresponding to the dynamics of a superpositional ``active'' state, $\sum_{ij}C_{ij}|00ij\rangle$. The second sum describes the dynamics of all the remaining 240 density matrix elements. These matrix elements should not change significantly during the dynamical process. Here we use a ``single-index'' decimal notation for each  four-spin state,
$$
|n\rangle=|psij\rangle,\quad n=j+2i+2^2s+2^3p.\eqno(34)
$$
The initial conditions for the matrix elements $b_{n,p}$ are: $b_{5,5}=b_{6,6}=b_{7,7}=b_{8,8}=1/2$, $b_{4,4}=b_{9,9}=b_{10,10}=b_{11,11}=-1/2$, $b_{12,12}=-1$,
$b_{13,13}=b_{14,14}=b_{15,15}=0$, $b_{n,p}=0$ for $n\not=p$. Our numerical calculations confirmed that single-pulse
 implementation of the CN gate performs well if the frequencies, $\omega_k$,  are well separated and the values of the Ising constants are not too small. For example, for the values of parameters,
$$
\omega_k=\omega_0+100\times k,\quad J_{kn}=J=10,\quad \Omega_k=\Omega=0.1, \eqno(35)
$$
we have demonstrated the ``complementary'' CN gate which drives the target spin if the control spin is in the ground state. For the initial superpositional state (30), after the action of the complementary CN gate, one creates the following wave function for the effective pure two-spin system of two ``active'' qubits,
$$
\Psi_1=i\sqrt{0.2}|00\rangle+i\sqrt{0.3}|01\rangle+{{1}\over{\sqrt{3}}}|10\rangle+{{1}\over{\sqrt{6}}}|11\rangle.\eqno(36)
$$
It is easy to check that the reduced density matrix, $r$, corresponding to the initial state (30) has the form:
$$
r_0=\left(\matrix{
0.3000&0.2449&0.3162&0.2236\cr
0.2449&0.2000&0.2582&0.1826\cr
0.3162&0.2582&0.3333&0.2357\cr
0.2236&0.1826&0.2357&0.1666\cr
}\right).\eqno(37)
$$
The density matrix corresponding to the wave function (36) is,
$$
r_1=\left(\matrix{
0.2000&0.2449&i0.2582&i0.1826\cr
0.2449&0.3000&i0.3162&i0.2236\cr
-i0.2582&-i0.3162&0.3333&0.2357\cr
-i0.1826&-i0.2236&0.2357&0.1666\cr
}\right).\eqno(38)
$$
The numerical solution of the equations of motion corresponding to the density matrix, $\rho_\Delta(t)$, in (33) showed  less than 0.5\% deviations of $r_{n,p}$ from the values in (38). With the same accuracy, the coefficients $b_{n,p}(t)$ in (33) do not change in time under the action of the $\pi$-pulse.

Next, we have studied numerically the range of parameters in which non-resonant effects remain small (with about 1\% deviations from the dynamics corresponding to the CN gate) \cite{b8}. According to our calculations, non-resonant effects do not destroy the single-pulse CN gate when $\Delta\omega/\Omega>300$, where $\Delta\omega=\omega_{k+1}-\omega_k$. This  is valid for a wide range of the values of the Ising constant of interaction: $3<J/\Omega<100$. 
An analysis of a single-pulse CN gate was also performed in \cite{c3} where it
was called the ``Pound-Overhauser implementation of the CN gate'' after the Pound-Overhauser double resonance in the NMR spectroscopy (on-transition excitation). 

To conclude this section, we mention related topics associated with a CN gate in Ising spin systems. The influence of small deviations from resonance and effects of ``noise'' have been studied in \cite{b4}. Two-pulse implementations of the CN gate have been considered in \cite{c4,c5}. In this case, the first short pulse induces a $\pi/2$ rotation of the target spin around the $x$-axis of the rotating reference frame. The second pulse induces a similar rotation around the $y$-axis. The delay time between the two pulses is $\pi/2J$. The action of the second pulse depends on the state of the control spin. For example, in experiments \cite{c4}, the control proton spin has the resonant frequency $500$MHz; the target carbon spin has the resonant frequency $125$MHz; and the Ising constant is: $J/2\pi=108$Hz. The two $\pi/2$-pulse sequence is more complicated than a single $\pi$-pulse implementation of the CN gate but it takes less time (approximately $\pi/2J$). Another two-pulse implementation of the CN gate was called the ``Pound-Overhauser on-resonance implementation of the CN gate'' \cite{c3}. It consists of a long $x$-pulse (with a duration of $\pi/(\sqrt{2}J)$, and a Rabi frequency of $\Omega=J$) which is tuned to the frequency between the peaks of the NMR doublet for the target spin, and a short $\pi/2$-pulse ($y$-pulse) covering the doublet.\\ \ \\
{\bf V. Elimination of non-resonant excitation}\\ \ \\

The question arises about the opportunities to eliminate non-resonant effects using special sequences of resonant pulses. One well-known approach to remove ``parts'' of the Hamiltonian in nuclear spin systems is ``selective averaging''. This  is one of the basic approaches for high resolution NMR in solids. (See, for example, \cite{h1} and references therein.) It was suggested recently to use this approach for analog quantum computations \cite{cl}. Indeed, using averaging (a proper sequence of pulses) one can transform with some accuracy the initial Hamiltonian into the desired Hamiltonian to study the evolution of the desired  quantum  system. In \cite{cl} this was done for two qubits, but the method can probably be extended for a larger number of qubits. Problems which one faces using this method have been discussed already in the first papers \cite{w1,w2}. (See also \cite{h1}.)

To the best of our knowledge, no one has proposed eliminating non-resonant part of the Hamiltonian using averaging. Instead of averaging, we suggested in \cite{b5,b9} a different method (the  ``$2\pi k$-method''): a $\pi$- or $\pi/n$ ($n=2,3,..$)- pulse which drives the resonant spin must simultaneously be $2\pi k$-pulse for a non-resonant spin. In this case, the non-resonant spin will return to its initial unperturbed state at the end of the excited pulse.

Consider, for example, two spins with a small frequency difference, $\Delta\omega$. If the exciting pulse, e.g. a $\pi$-pulse, is tuned to the frequency of the first spin, we have,
$$
\Omega\tau=\pi,\eqno(39)
$$
where $\tau$ is the duration of the pulse and $\Omega$ is the Rabi frequency. For the non-resonant spin, the effective field in the rotating frame in  frequency units is,
$$
\omega_e=\sqrt{\Omega^2+\Delta\omega^2},\eqno(40)
$$
where we have assumed equal Rabi frequencies for both spins. 
If,
$$
\omega_e\tau=2\pi k,\quad k=1,2,...,\eqno(41)
$$
the non-resonant spin is not deflected by the pulse. From (39)-(41) we easily derive the required conditions for the Rabi frequency and for the duration of the $\pi$-pulse which allow one to eliminate undesirable non-resonant excitations \cite{b5}. (See also Chapter 22 in \cite{b9}.),
$$
\Omega={{|\Delta\omega|}\over{\sqrt{4k^2-1}}},\quad \tau=\pi/\Omega.\eqno(42)
$$

(We do not discuss here the phase effect for a $2\pi$-pulse: for a pure quantum state, a $2\pi$-pulse changes the phase of the state by $\pi$ and a $4\pi$-pulse returns the system into the initial quantum state. 
See, for example, \cite{b9}.) 
By manipulating $\Delta\omega$, one can eliminate non-resonant effects for two or more weakly interacting spins. Assume, for example, that the frequencies of spins are,
$$
\omega_0,~\omega_1=\omega_0+8\Omega,~\omega_2=\omega_0+16\Omega,~\omega_3=\omega_0+24\Omega,...,\omega_n=\omega_0+8n\Omega,~(n=1,2,...).\eqno(43)
$$
If one applies a resonant $\pi$-pulse with any of these frequencies, e.g., $\omega=\omega_2$, the frequency difference, $|\Delta\omega|$, takes the values,
$$
8\Omega,~16\Omega,~24\pi,..., 8n\Omega,~(n=1,2,...).\eqno(44)
$$
According to (40), (41), (44) the corresponding angles of rotation for the non-resonant spins are,
$$
\sqrt{\Omega^2+\Delta\omega^2}\times{{\pi}\over{\Omega}}=\pi\sqrt{\Bigg({{\Delta
\omega}\over{\Omega}}\Bigg)^2+1}\approx{{\pi|\Delta\omega}|\over{\Omega}}=8\pi,~16\pi,~24\pi,...,8n\pi,~(n=1,2,...).\eqno(45)
$$
It is clear from (45) that at the end of the $\pi$-pulse, all non-resonant spins will return to states which are very close to the initial states. 

For $\Omega=10^3s^{-1}$, the frequency difference is $\Delta\omega=8\Omega$ which corresponds to a change of the external magnetic field, $B$, for the neighboring protons by: $\Delta B\approx 2\times 10^{-5}$T. Choosing the distance between neighboring protons, $\Delta x=1$nm, we obtain an estimate for the required magnetic field gradient,
$$
{{\Delta B}\over{\Delta x}}\approx 3\times 10^{4}T/m. \eqno(46)
$$
This gradient is easily achievable, for example, in recent Magnetic Resonance Force Microscopy experiments \cite{christ}.

For a $\pi/n$-pulse ($n=2,3,...$), instead of (42) we have the generalized condition,
$$
\Omega={{|\Delta\omega|}\over{\sqrt{(2nk)^2-1}}},~(k=1,2,...).\eqno(47)
$$
It was suggested in \cite{c3} that the same $2\pi k$-method be used to provide an ``exact'' single-pulse CN gate in the two-spin system. If a control spin is in the ground state, the frequency of the transition for the target spin is $\omega_0+J$, where $\omega_0$ is the target spin frequency ignoring Ising interactions. If the control qubit is in the excited state, the frequency of the transition is $\omega_0-J$. So, in this case, the frequency difference in (42) is: $\Delta\omega=2J$. A pulse with frequency $\omega=\omega_0-J$ and with the Rabi frequency and duration from (42) represents a resonant $\pi$-pulse if the control spin is in the excited state, and a $2\pi k$-pulse if the control spin is the the ground state. In the second case, the target spin returns to its initial state. Choosing $k=1$, one can reduce the duration of a single-pulse CN gate to: $\sqrt{3}\pi/2J$.\\ \ \\
{\bf Conclusions}\\ \ \\
In this paper we discuss problems related to the dynamics of quantum computation. We presented the simplest demonstration of the dynamics of Shor's algorithm using only four qubits. In this example, we showed that  fast oscillations can destroy the desired quantum interference. We also demonstrated how this effect can be routinely eliminated using resonant pulses. We also  considered influence of non-resonant effects on quantum computation. We have shown that for well-separated frequencies, a non-resonant excitation does not prevent a single-pulse implementation of logic gates for both pure quantum states and for an ensemble of quantum systems. We have found the range of parameters where non-resonant effects remain small. Finally, we have described a ``$2\pi k$ -method'' which allows one to drive resonant spins without deflecting of non-resonant spins. Future studies of dynamical problems related to quantum computation will be directed to many-qubit systems and the influence of noise in real systems. \\ \ \\
{\bf Acknowledgments}\\ \ \\
This work  was supported by the Department of Energy under contract W-7405-ENG-36, and by the National Security Agency.
\newpage
\end{document}